\documentstyle[12pt]{article}
\begin{document}
\bibliographystyle{plain}

\begin{center} {\bf QCD Sum-Rule Consistency of Lowest-Lying $q
\bar{q}$ Scalar Resonances}
\end{center}

\bigskip

\begin{center}V. Elias and A. H. Fariborz$^*$ \\ Department of Applied
Mathematics \\ The University of Western Ontario \\ London, Ontario
N6A 5B7 \\ Canada
\end{center}

\bigskip

\begin{center} and
\end{center}

\bigskip 

\begin{center}Fang Shi and T. G. Steele \\ Department of Physics and
Engineering Physics \\ University of Saskatchewan \\ Saskatoon,
Saskatchewan  S7N 5C6 \\ Canada
\end{center}

\bigskip

\noindent $^*$ Current Address:  Department of Physics, Syracuse
University, Syracuse, New York 13244-1130, USA.

\newpage

\begin{abstract}

      We investigate lowest-lying scalar meson properties predicted from QCD Laplace sum
rules based upon isovector and isoscalar non-strange $\bar{q}q$ currents.  The hadronic content
of these sum rules incorporates deviations from the narrow resonance approximation anticipated
from physical resonance widths.  The field theoretical content of these sum rules incorporates
purely-perturbative QCD contributions to three-loop order, the direct single-instanton contribution
in the instanton liquid model, and leading contributions from QCD-vacuum condensates.  In the
isovector channel, the results we obtain are compatible with a$_0$(1450) being the lowest-lying
$q\bar{q}$ resonance, and are indicative of a non-$q\bar{q}$ interpretation for a$_0$(980).  In
the isoscalar channel, the results we obtain are compatible with the lowest lying
$q\bar{q}$
resonance being f$_0$(980) or a state somewhat lighter than f$_0$(980) whose width is less than
half of its mass.  The dilaton scenario for such a narrower $\sigma$-resonance is discussed in
detail, and is found compatible with sum rule predictions for the resonance coupling only if the
anomalous gluon-field portion of $\Theta_\mu^\mu$ dominates the matrix element
$<\sigma|\Theta_\mu^\mu|0>$. A linear sigma-model interpretation of the lowest-lying
resonance's coupling, when compared to the coupling predicted by sum rules, is indicative of a
renormalization-group invariant light-quark mass between 4 and 6 MeV.  
\end{abstract}

\section{Introduction: Status of the Lowest-Lying Scalar
Resonances}

At present, there is a great deal of confusion concerning both the identity and
interpretation of the lowest lying I = 0 and I = 1 scalar resonances, specifically the four states
denoted in the Particle Data Guide by $f_0$(400-1200), $f_0$(980),
$a_0$(980), and $a_0$(1450).  The
nearness of the $f_0$(980) and $a_0$(980) to a $K \bar{K}$ threshold has led to a widely held
interpretation of
these states as isopartner $K \bar{K}$-molecules [1,2,3], as opposed to light
$q \bar{q}$-resonances (linear
combinations of $u \bar{u}$ and $d \bar{d}$ states).  However, the assumption that these states
are isopartners
and/or $K \bar{K}$-molecules have both been subject to recent scrutiny.  In particular, 
Morgan and Pennington [4] have disputed the $K \bar{K}$
interpretation of $f_0$ (980).  An analysis
using the J\"ulich model for $\pi\pi$ scattering [5] {\it is} compatible with a
$K \bar{K}$ interpretation of $f_0$(980),
but sees $a_0$(980) as a dynamical threshold effect, as opposed to a true resonance state.  An
even
more recent analysis of OPAL data [6]
supports the consistency
of a $q \bar{q}$ interpretation of
$f_0(980)$.

Theory is similarly ambivalent regarding the $f_0$(980) and $a_0$(980) scalar resonance states. 
A QCD sum rule analysis [7] based upon correlation-function currents chosen to project out
$K \bar{K}$-molecule states concludes that $f_0$(1500) and $f_0$(1710) are better candidates
for such multiquark
states than either $f_0$(980) or $a_0$(980).  Moreover, a very recent coupled 
channel analysis of
$\pi \pi$ scattering [8]
suggests that the $f_0$(980) state may really be two distinct S-matrix poles
(see also [4]), one corresponding
to a $K \bar{K}$ molecule and the other perhaps corresponding to a light $q
\bar{q}$ state.

Indeed, the determination of the lowest lying $q \bar{q}$ I = 0 and I = 1 states is of genuine
value as a test of our present understanding of QCD, particularly its nonperturbative content. 
Such lowest lying states, when first compared with QCD via sum rule methods
[9], were necessarily
found to be degenerate, as purely-perturbative and QCD-vacuum condensate contributions to
scalar-current correlation functions cannot distinguish between I = 0 and I = 1 channels. Of
course, this result was first seen to account for the degeneracy of
$f_0$(980) and $a_0$(980) as lowest-lying $q \bar{q}$ isopartners [9,10].  As has been
emphasized repeatedly over the last twenty years,
however, both scalar and pseudoscalar channels exhibit significant sensitivity not only to
nonperturbative field-theoretical effects with infinite correlation length (QCD-vacuum
condensates), but also to instanton effects, the nonperturbative content of the QCD vacuum
characterized by finite correlation lengths [11,12,13,14,15,16].  The instanton
component of the QCD vacuum is known to distinguish between I = 0 and I = 1 scalar (and
pseudoscalar) states [12,13].  Such a distinction is, of course, quite evident in the pseudoscalar
channel's large $\pi - \eta$ mass difference, though an understanding of the pseudoscalar I = 1
channel
necessarily must take into account the first pion-excitation state because of the near-masslessness
(and concomitantly reduced sum-rule contribution) of the pion [17,18,19].  Similarly, the
existence of instanton solutions in QCD {\it necessarily imposes the
theoretical expectation that a similar split occur between I = 0 and I = 1
$q \bar{q}$ scalar resonance
states, with the I = 0 state substantially lighter than its I = 1
isopartner} [12,13]. In this regard,
scalar meson spectroscopy is a genuine test of QCD.

Recent activity [20,21,22,23] in re-analyzing old $\pi \pi$ and $\pi$N scattering data has
led to the reinstatement of a lowest-lying I = 0 scalar resonance that is distinct
from the $f_0$(980),
conservatively labelled by the 1996 Particle Data Group [1] as $f_0$(400-1200). Perhaps of equal
importance, assuming $a_0$(980) and $f_0$(980) really are either $K
\bar{K}$ isopartners or other non-$q \bar{q}$ exotica, is the Crystal Barrel 
Collaboration's recent confirmation of an $a_0$(1450) I = 1 scalar
resonance state [24].  If this isovector state is identified as the lowest-lying
I = 1 {\it $q \bar{q}$-scalar}
resonance (lattice simulations have shown glueballs below 1600 MeV to be unlikely), it is
important to determine whether the identification of its I = 0 isopartner with any portion of the
$f_0$(400-1200) mass range (particularly the high end [4,25]) is compatible with the
instanton-generated mass difference anticipated from QCD.

Moreover, the $f_0$(400-1200) has been widely interpreted to be the
$\sigma$-particle signature of
chiral symmetry breaking anticipated from NambuþJona-Lasinio (NJL) dynamics
[26], the
linear sigma-model (L$\sigma$M) spectrum [27], as well as in models for
$q \bar{q}$ scattering in an instanton
background [12,13] and in one boson exchange models of the nucleon-nucleon potential
[28]. 
Such a $\sigma$-particle, however, does {\it not} characterize the
{\it nonlinear} sigma model (NL$\sigma$M); indeed,
the empirical absence of a credible $\sigma$ prior to 1996 has provided impetus for the
development
of NL$\sigma$M ideas into a chiral perturbation theory framework.  Clearly a clarification of the
properties, or even the existence [29], of a light $\sigma$-resonance is required to distinguish
between
L$\sigma$M and NL$\sigma$M alternatives for effective theories of low-energy hadron physics.

We choose here to distinguish, somewhat arbitrarily, four different alternatives for the
$f_0$(400-1200) resonance that each have some empirical support:

\begin{enumerate}
\item The resonance is both very light ($m_\sigma \stackrel{<}{_\sim}$ 500 MeV) and very
broad
($\Gamma_\sigma \stackrel{>}{_\sim}$ 500 MeV), as suggested by DM2 data
[30] and by
T\"ornqvist and Roos's analysis of $\pi \pi$ scattering [23].  It
must be recognized, however, that such a resonance may be generated
dynamically [8] and is not necessarily a $q\bar{q}$ state.

\item The resonance is $\sigma$-like in mass (500-700 MeV) but substantially narrower in
     width ($\Gamma_\sigma \stackrel{<}{_\sim} m_\sigma /2$), consistent with parameter ranges
extracted by 
     Svec [20], S. Ishida et al [21], and Harada, Sannino, and Schechter
[22].

\item The lowest I = 0 $q \bar{q}$ scalar is to be identified with a
$q \bar{q}$ pole at (or perhaps masked
     by) the $f_0$(980) resonance [8], with a narrow ($\Gamma_\sigma
\stackrel{<}{_\sim}$ 150 MeV) width.

\item The lowest I = 0 $q \bar{q}$ scalar resonance is a very broad structure in the
$\pi \pi$ scattering amplitude characterized by a mass comparable to
[4] or substantially above [25] that of the $f_0$ (980).
\end{enumerate}

There is, of course, some blurring of the boundaries between these alternatives,
particularly Alternatives 2 and 3.  Svec [31] has recently reported a single-pole fit of
$\pi$-N(polarized)
scattering data leading to a mass (775 $\pm$ 17 MeV) somewhat larger than
NJL-L$\sigma$M expectations,
and a width (147 $\pm$ 33 MeV) substantially narrower than those already quoted in support of
Alternative 2.  Moreover, Harada, Sannino, and Schechter have demonstrated
[32] how a very
light, very broad state [Alternative 1] can transmute into a heavier, narrower state [Alternative
2] when $\rho$ exchanges are taken into account.  From a theoretical point of view, Alternatives
1
and 2 support the existence of a $\sigma$-particle, though straightforward
L$\sigma$M expectations would
favour the mass range of Alternative 2 and the broad width of
Alternative 1 [33,34].  Theoretical arguments for a
narrower $\sigma$-particle more fully consistent with Alternative 2 have been advanced through
identifying the $\sigma$ with the near-Goldstone particle of dilatation symmetry in the strong
coupling
limit [35,36].  Such a dilaton would be expected to have a width similar to that of the
$\rho$-meson, corresponding to cancellation of an enhancement factor of 9/2 in a naive
calculation of
the width [33] against an anticipated suppression factor
$1/d_\sigma^2 \approx 1/4$, where $d_\sigma$ is the anomalous
mass dimension in the strong coupling limit. \footnote{Note that
$f_\pi$ appearing in eq. (13) of [35] should be understood to be 131
MeV, not 93 MeV, so that the $d_\sigma = 1$ prediction of
$\Gamma_\sigma$ coincides with the L$\sigma$M-equivalent prediction
in [34].  This point has been verified through personal communciation
with V. A. Miransky.}

In the present manuscript, we employ QCD Laplace sum-rules as a technique particularly
well-suited to relate the field-theoretical content of QCD to lowest-lying resonance properties
[14]. We use sum-rule methodology specifically to address the following questions:

\bigskip

\noindent{\it Which, if any, of the four alternatives discussed above for the lightest I = 0
$q \bar{q}$ scalar are supported by QCD sum rules?}  In particular, do QCD
sum rules rule
out either of the alternatives that are consistent with an
NJL/L$\sigma$M $\sigma$-like object?  If the existence of a
$\sigma$ is consistent with QCD sum rules, is such a $\sigma$ a broad
L$\sigma$M object, or a narrower strong-coupling dilaton?

\bigskip

\noindent{\it What is the mass range for the lightest I = 1 $q \bar{q}$ scalar
resonance?}  In particular, can we rule out all but exotic interpretations 
of the $a_0$(980), and does there exist sum-rule support for the recently
confirmed $a_0$(1450) being the lowest-lying $q \bar{q}$ object in this
channel? 

\bigskip

      In the section that follows, we present the sum rule methodology necessary to address
these questions.  Specifically, we show how nonzero resonance widths can be incorporated into
the hadronic contribution to QCD Laplace sum rules, which are argued to be particularly
appropriate for studying lowest-lying resonance properties.  We also demonstrate explicitly how
a lowest-lying resonance's nonzero width elevates a sum rule determination of that resonance's
mass.

      In Section 3 we present the field-theoretical content of appropriate scalar current sum
rules.  We discuss the sum rule contribution arising from the 3-loop order purely-perturbative
QCD contributions to the scalar current correlation function.  We also present nonperturbative
sum rule contributions arising from QCD-vacuum condensates and direct single-instanton
contributions to the I = 0 and I = 1 scalar current correlation functions.

      In Section 4, we utilize the results of the preceding two sections to obtain a sum-rule
determination of the masses of lowest-lying scalar resonances.  Stability curves are generated
leading to estimates of such masses for a given choice of width and the continuum threshold
above which perturbative QCD and hadronic physics are assumed to coincide.  Detailed
comparison is made with earlier sum-rule generated stability curves [9], showing how the
separate incorporation of renormalization-group improvement, 3-loop perturbative effects, nonzero
widths, and the contribution of instantons individually affect such curves.

      In Section 5, we examine the isoscalar channel in further detail by obtaining values of the
mass, width, continuum threshold, and coupling of the lowest-lying
$q\bar{q}$ resonance via a weighted
least-squares fit to the overall Borel-parameter dependence of the first Laplace sum rule.  A
relationship between the anticipated resonance coupling and a phenomenologically estimable
matrix element is developed in Appendix A.  In Section 5 this relationship is utilized to obtain
an estimate of the light quark mass.  This relationship is also used
to assess the sum-rule consistency of a dilaton interpretation for
the lowest-lying resonance.

      Finally in Section 6 we present our conclusions concerning the questions we have raised
above. We assess the compatibility of sum rule predictions for the lowest-lying non-exotic I=1
resonance
with a$_0$(980) and a$_0$(1450). We
examine in detail the four alternatives presented above for interpreting the I = 0 scalar resonance
spectrum and argue that Alternatives 1 and 4 appear to
be unsupported by a sum-rule based analysis of the lowest-lying
$q\bar{q}$ state. We also discuss how our conclusions are affected
by possible sum rule contamination
from higher resonances.

\newpage

\section{Sum-Rule Methodology and Lowest-Lying $q \bar{q}$-Scalar Resonances of
Nonzero Width}

In the narrow resonance approximation, subcontinuum resonance contributions to the 
light-quark scalar-current correlation function,\footnote{We have
normalized I=0 (+) and I = 1 (-) scalar currents in (2) so as to
facilitate comparison with ref. [9].}

\begin{equation}
\Pi(p^2) = i \int d^4 x e^{ip \cdot x} <0| T j(x) j(0) |0>,
\end{equation}

\begin{equation}
j(x) \equiv [\bar{u} (x) u (x) \pm  \bar{d} (x) d(x)] / 2,
\end{equation}

are proportional to a sum of delta functions:

\begin{eqnarray}
(Im \Pi(s))_{res.} & = & Im \sum_r \left[ \frac{-g_r}{(s-m_r^2) + i m_r
\Gamma_r} \right] \nonumber \\
& = & \sum_r \left[ \frac{g_rm_r\Gamma_r}{(s-m_r^2)^2 + m_r^2 \Gamma_r^2}
\right] \nonumber \\
& & \begin{array} {c} {} \\ \longrightarrow \\ \Gamma_r \rightarrow 0
\end{array} \sum_r \pi g_r
\delta (s-m_r^2).
\end{eqnarray}

The coupling coefficient $g_r$ is proportional to $m_r^2$. However, the constant of
proportionality is
expected to be much larger for $q \bar{q}$ resonances [i.e. resonances that couple directly to the 
field-theoretical operators in the scalar current (2)] than for exotic resonances
[37].  It is for
precisely this reason that sum-rule searches for non-$q \bar{q}$ scalar resonance states, 
such as $K \bar{K}$
molecules [7] or glueballs [37,38], utilize correlation functions based on appropriate 
$K \bar{K}$- or gluonic-currents that couple directly to such hadronic exotica.

Laplace sum rules $R_k(\tau)$ for the scalar current correlation function are particularly
sensitive to the lowest-lying $q \bar{q}$ resonance of a given
isostructure:

\begin{eqnarray}
R_k(\tau) & \equiv & \frac{1}{\pi} \int_0^\infty s^k Im [\Pi(s)] e^{-s
\tau} ds \nonumber \\
& = & \sum_r  g_r m_r^{2k} e^{-m_r^2 \tau} \Theta (s_0 - m_r^2)
\nonumber \\
& + & \frac{1}{\pi} \int_{s_0}^\infty s^k Im [(\Pi(s))_{pert.} ] e^{-s
\tau} ds.
\end{eqnarray}

\noindent The summation over resonances in (4) clearly follows from the final line of (3).  The
remaining
integral in (4) reflects the anticipated duality [39] between purely-perturbative QCD and
hadronic physics above some appropriately chosen continuum threshold
$s > s_0$.  As is evident
from (4), higher-mass resonances are either absorbed into the
continuum $(m_r^2 > s_0)$, or if subcontinuum $(m_r^2 < s_0)$,
are exponentially suppressed relative to low-mass resonances.
Consequently, Laplace sum rules are well-suited for determining
properties of the lowest-lying resonance in a given channel. The
subcontinuum resonance contribution $R_k(\tau,s_0)$ to the $k^{th}$ Laplace sum 
rule, defined as 

\begin{equation}
R_k(\tau, s_0) \equiv R_k(\tau) - \frac{1}{\pi} \int_{s_0}^\infty
s^k Im [(\Pi(s))_{pert.}] e^{-s \tau} ds,
\end{equation}

\noindent is clearly seen from (4) to satisfy the inequality

\begin{equation}
\frac{R_{k+1}(\tau, s_0)}{R_k(\tau, s_0)} \geq m_\ell^2,
\end{equation}

\noindent where $m_\ell$ is the mass of the lowest-lying resonance in a given channel.  QCD
sum-rule methodology for a given resonance channel generally involves obtaining an estimate
of the mass $m_\ell$ via minimization of the field-theoretical content of the left-hand side of (6)
with
respect to the Borel parameter $\tau$ [14,40].  In practice, the ratio utilized is
$R_1/R_0$, so as to avoid
methodologically inconvenient enhancement of continuum and higher-mass 
resonance contributions, as well as heightened dependence on
(unknown) higher dimensional condensates, 
through the factors $m_r^{2k}$ appearing in (4).

The field theoretical content of the first two Laplace sum rules
$R_{0,1}(\tau,s_0)$ will be discussed
in the section that follows.  However, it is important to recognize that (6) requires significant
modification if the lowest-lying resonance is broad.  If one does not invoke the narrow resonance
approximation in the final line of (3), but instead assumes a Breit-Wigner (or modified Breit-
Wigner [1]) shape, one finds that

\begin{eqnarray}
R_k(\tau, s_0) & \cong &  \frac{1}{\pi} \sum_r^{m_r^2 < s_0} \int_0^\infty
\frac{g_r m_r\Gamma_r}{(s-m_r^2)^2 + m_r^2\Gamma_r^2} s^k e^{-s \tau}
ds \nonumber \\
& = & \sum_r g_r W_k (m_r, \Gamma_r, \tau)m_r^{2k} e^{-m_r^2 \tau}
\Theta(s_0 - m_r^2),
\end{eqnarray}

\noindent with the weighting functions $W_k$ demonstrably related to the narrow resonance limit
(4) via

\begin{equation}
\lim_{\Gamma_r \rightarrow 0} W_k (m_r, \Gamma_r, \tau) = 1.
\end{equation}

\noindent The net effect of such weighting functions on the lowest-lying resonance contribution
to (6) is
to replace $m_\ell^2$ with $m_\ell^2  W_{k+1}/W_k$. If the
lowest-lying resonance is the dominant subcontinuum resonance in a
a given channel, then
$m_\ell$ can be extracted from the lowest-lying $(\ell)$
resonance contribution to $R_1/R_0$ as follows [19,41]:

\begin{equation}
\frac{W_0(m_\ell, \Gamma_\ell, \tau)}{W_1(m_\ell, \Gamma_\ell, \tau)}
\left( \frac{ R_1(\tau, s_0)}{R_0(\tau, s_0)} \right)_\ell =
m_\ell^2.
\end{equation}

For a given choice of $\Gamma_\ell$ and $s_0$, one can use
field-theoretical expressions for $R_{0,1}(\tau,s_0)$, including both
perturbative QCD and nonperturbative QCD contributions of infinite
and finite correlation lengths [Section 3], to obtain from (9) a
self-consistent minimizing value of $m_\ell$.  This procedure
constitutes the methodological foundation for the results we obtain
in Section 4.

The weighting functions $W_{0,1}$ can be derived from a Breit-Wigner
resonance shape by expressing that shape as a Riemann sum of
unit-area pulses $P_{m_r}$ centred at $s = m_r^2$ [19]:

\begin{equation}
P_M [s, \Gamma] \equiv \frac{1}{2M\Gamma} [ \Theta (s - M^2 + M
\Gamma) - \Theta (s - M^2 - M \Gamma)],
\end{equation}

\begin{equation}
\frac{M \Gamma}{(s-M^2)^2 + M^2 \Gamma^2} = \lim_{n \rightarrow
\infty} \frac{2}{n} \sum_{j=1}^n \sqrt{ \frac{n-j+f}{j-f}} P_M \left[
s, \sqrt{\frac{n-j+f}{j-f}} \; \Gamma \right],
\end{equation}
where $f$ is any arbitrarily chosen constant between $0$ and $1$. If one
approximates the resonance shape via (11) by truncating $n$ to some
finite number of pulses, then the approximation (unlike the $n
\rightarrow \infty$ limit) becomes sensitive to the choice of $f$.
The value for $f$ may be chosen to ensure that the
area under the truncated sum is equal to the area under the ``true''
resonance shape

\begin{equation}
\int_{-\infty}^\infty \frac{M \Gamma}{(s-M^2)^2 + M^2 \Gamma^2} ds =
\pi.
\end{equation}
In an $n=4$ approximation, for example, one obtains an area of $\pi$
by choosing $f = 0.70$.  One finds for this four-pulse approximation
[Fig. 1] that [19]

\begin{eqnarray}
W_k[M, \Gamma, \tau] & \cong & 0.5589 \Delta_k (m, 3.5119 \; \Gamma,
\tau) \nonumber \\
& + & 0.2294 \Delta_k (M, 1.4412  \; \Gamma, \tau) \nonumber \\
& + & 0.1368 \Delta_k (M, 0.8597  \; \Gamma, \tau) \nonumber \\
& + & 0.0733 \Delta_k (M, 0.4606  \; \Gamma, \tau), 
\end{eqnarray}
where
\begin{equation}
\Delta_k (M, \Gamma, \tau) M^{2k} e^{-M^2 \tau} \equiv \int
_{-\infty}^\infty P_M (s, \Gamma) s^k e^{-s \tau} ds.
\end{equation}
In particular,
\begin{equation}
\Delta_0 (M, \Gamma, \tau) = \frac{sin h(M \Gamma \tau)}{M \Gamma
\tau},
\end{equation}
and
\begin{equation}
\Delta_1 (M, \Gamma, \tau) = \frac{sin h (M \Gamma \tau)}{M \Gamma
\tau} \left[ 1 + \frac{1}{M^2 \tau} \right] - \frac{cos h (M \Gamma
\tau)}{M^2 \tau}.
\end{equation}
As $\Gamma \rightarrow 0$, $\Delta_k \rightarrow 1$.  Nevertheless,
it is easy to show for small values of $\Gamma$ that [41]

\begin{equation}
\frac{\Delta_0(M, \Gamma, \tau)}{\Delta_1 (M, \Gamma, \tau)}
= 1 + \Gamma^2 \tau /
3 + O(\Gamma^4)
\end{equation}
in which case we find from (13) that $W_0/W_1 > 1$.

It is evident from a comparison of (6) and (9) that the introduction
of a small nonzero width necessarily increases a sum rule driven
estimate of $m_\ell$ [41].  Specifically, if we regard (9) as a
constraint implicitly defining a function $m_\ell(\tau, \Gamma_\ell,
s_0)$, the result $W_0/W_1 > 1$ for nonzero $\Gamma$ implies that

\begin{equation}
m_\ell (\tau, \Gamma_\ell, s_0) > m_\ell (\tau, 0, s_0).
\end{equation} 
Such behaviour is evident from the analysis carried out in Section 4.

\newpage 

\section{Field Theoretical Contributions to Scalar-Current
Sum Rules}

In this section, we seek to identify purely-perturbative and nonperturbative QCD
contributions to $R_0(\tau)$ and $R_1(\tau)$, as defined by (4).  The three-loop expression for
the scalar-current [as defined by (2)] correlation function (1) in perturbative QCD 
$(n_f = 3; \;  s = p^2 = -Q^2)$
is given by [42,43]

\begin{eqnarray}
\Pi_{pert}(s) & = & \frac{3}{16 \pi^2} (-s) \ell n \left(
\frac{-s}{\mu^2}\right) \left\{1 + \left( \frac{\alpha_s}{\pi}\right)
\left[ \frac{17}{3} -\ell n \left( \frac{-s}{\mu^2}\right) \right]
\right. \nonumber \\
& + & \left.  \left( \frac{\alpha_s}{\pi} \right)^2 \left[ 45.846 -
\frac{95}{6}\ell n \left( \frac{-s}{\mu^2}\right) + \frac{17}{12}
\ell n^2 \left( \frac{-s}{\mu^2} \right) \right] \right\}
\end{eqnarray}

\noindent The imaginary part of (19), needed for the integrand of (4), is defined consistent with
dispersion-relation conventions:

\begin{equation}
2i Im \Pi (s) = \Pi (s+i\epsilon) - \Pi (s - i \epsilon),
\end{equation}

\noindent [e.g. $Im(\ell n(-s/\mu^2)) = -\pi]$, in which case one finds from (19) that

\begin{eqnarray}
Im \left( \Pi_{pert} (s) \right) & = & \frac{3s}{16\pi} \left\{1 +
\left( \frac{\alpha_s}{\pi} \right) \left[ \frac{17}{3} - 2 \ell n
\left( \frac{s}{\mu^2}\right) \right] \right. \nonumber \\
& + & \left. \left( \frac{\alpha_s}{\pi} \right)^2 \left[ 31.864 -
\frac{95}{3} \ell n \left( \frac{s}{\mu^2}\right) + \frac{17}{4} \ell
n^2 \left( \frac{s}{\mu^2} \right) \right] \right\}
\end{eqnarray}

\noindent We choose to set the renormalization scale via the Borel parameter
by setting $\mu^2 = 1/\tau$.  Upon
substituting (21) into (4), we obtain from (5) the following purely-perturbative 
contribution to $R_0(\tau,s_0)$:

\begin{eqnarray}
\left[ R_0(\tau, s_0) \right]_{pert} \nonumber \\
& = & \frac{3}{16 \pi^2 \tau^2} \left\{ \left[ 1 - (1 + s_0 \tau)
e^{-s_0 \tau} \right] \left[ 1 + \left( \frac{\alpha_s}{\pi} \right)
\frac{17}{3} + \left( \frac{ \alpha_s}{\pi} \right)^2 31.864 \right]
\right. \nonumber \\
& - & \left( \frac{\alpha_s}{\pi} \right) \left[ 2 + \frac{95}{3}
\left( \frac{\alpha_s}{\pi} \right) \right] \int_0^{s_0 \tau} w \; \ell n (w)
e^{-w} dw \nonumber \\
& + & \left. \left( \frac{ \alpha_s}{\pi}\right)^2 \frac{17}{4} \int_0^{s_0
\tau} w \left[ \ell n (w) \right]^2 e^{-w} dw \right\}
\end{eqnarray}

\noindent As is evident from (4) and (5), the subsequent sum rule
$R_1(\tau,s_0)$ can be obtained from (22) by
explicit differentiation with respect to the Borel parameter $\tau$.

     QCD-vacuum condensate contributions to the scalar-current correlation function given by
(1) and (2) are found via operator product methods to be [9,14,44]:

\begin{eqnarray}
\left( \Pi (s = -Q^2) \right)_{cond} & = & \left( \frac{3}{2Q^2} -
\frac{m_q^2}{Q^4} \right) <m_q \bar{q}q> \nonumber \\
& + & \left( \frac{1}{16 \pi Q^2} + \frac{7m_q^2}{24 \pi Q^4} -
\frac{m_q^2}{4 \pi Q^4} \ell n \left( \frac{Q^2}{\mu^2} \right)
\right) < \alpha_s G^2 > \nonumber \\
& + & \left( \frac{m_q}{2Q^4} - \frac{3 m_q^3}{2 Q^6} \right) <
\bar{q} G \cdot \sigma q > \nonumber \\
& + & \left( \frac{ 27 m_q^2}{48 \pi Q^6} - \frac{m_q^2}{4 \pi Q^6}
\ell n \left( \frac{Q^2}{\mu^2} \right) \right)  < \alpha_s G^3>
\nonumber \\
& + & O \left( m_q^4 \right) - \left( \frac{88 \pi}{27 Q^4} + O
\left( m_q^2 \right) \right) < \alpha_s ( \bar{q}q)^2 > ,
\end{eqnarray}

\noindent where $m_q = m_u = m_d$ in the SU(2)-flavour symmetry limit.  The final term 
in (23) is obtained via
the assumed vacuum saturation of several contributing dimension-6 operators [9].

     Standard dispersion-relationship arguments link the expression (4) for 
$R_0(\tau)$ with $d \Pi/dQ^2$:

\begin{equation}
- \frac{d \Pi}{d Q^2} = \frac{1}{\pi} \int_0^\infty \frac{Im
\Pi(s)}{(s + Q^2)^2} ds
\end{equation}

\noindent Equation (24) may be expressed as a Laplace transform (with respect 
to the variable $Q^2$) of the function $R_0(\tau)$.  Since

\begin{equation}
\frac{1}{(s+Q^2)^2} = \int_0^\infty (\tau e^{-s \tau}) e^{-Q^2 \tau}
d \tau \equiv {\cal L} (\tau e^{-s \tau}),
\end{equation}

we see from (24) and (4) that

\begin{equation}
{\cal L}^{-1} \left( - \frac{d \Pi}{dQ^2} \right) = \frac{1}{\pi}
\int_0^\infty \tau e^{-s \tau} Im \Pi(s) ds = \tau R_0 (\tau).
\end{equation}

\noindent We note from the Laplace transform definition in (25) that
${\cal L} [\tau] = 1/Q^4,  {\cal L}[1] = 1/Q^2$, and we
find through substitution of (23) into the left-hand side of (26) 
that the leading-order condensate contribution to $R_0$ is given by [9]

\begin{eqnarray}
\left[ R_0 (\tau) \right]_{cond} & = & \frac{3}{2} < m_q \bar{q}q> +
\frac{1}{16 \pi} < \alpha_s G^2> \nonumber \\
& - & \frac{88 \pi}{27} < \alpha_s (\bar{q}q)^2 > \tau \nonumber \\
& + & {\cal O}(m_q).
\end{eqnarray}

\noindent We do not include $<m_q \bar{q}q>$ in the order $m_q$
terms, as this condensate's magnitude $(= -f_\pi^2 m_\pi^2 / 4)$
is independent of the quark mass [14,45]. 

As discussed in Section 1, both scalar and pseudoscalar correlation functions are also
sensitive to nonperturbative contributions of finite correlation length, corresponding to the
instanton structure of the QCD vacuum.  The direct single-instanton contribution to the I = 1
{\it pseudoscalar} sum rule $R_0^\pi$ [16,46],

\begin{equation}
\left[ R_0 (\tau) \right]_{inst.}^\pi = \frac{3 \rho^2}{16 \pi^2
\tau^3} e^{-\rho^2 / 2 \tau} \left[ K_0 \left( \frac{\rho^2}{2 \tau}
\right) + K_1 \left( \frac{\rho^2}{2 \tau} \right) \right],
\end{equation}

\noindent is constructed via (4) from a correlation function based on the pseudoscalar current

\begin{equation}
j^p(x) \equiv i \left[ \bar{u}(x) \gamma_5 u (x) - \bar{d} (x)
\gamma_5 d (x) \right] / 2.
\end{equation}

\noindent This direct single-instanton correlator

\begin{equation}
\left( \Pi (p^2) \right)_{inst}^p \equiv i \int d^4 x e^{i p \cdot
x} <0| T \left[ j^p (x) j^p (0) \right]_{inst}|0>
\end{equation}

\noindent can be related to the direct single-instanton contribution to the I = 1 pseudoscalar
channel of the
quark-antiquark scattering amplitude [12] simply by tying the external fermion lines together [Fig.
2] to form the vacuum-polarization loop corresponding to (30).  Consequently, useful properties
of the single-instanton contribution to $q \bar{q}$ scattering are also applicable to (30). 
Specifically, the
I = 1 pseudoscalar $q \bar{q}$ scattering amplitude in a single-instanton background is
equivalent to the
I = 0 scalar $q \bar{q}$ amplitude [12,13], the result of compensating sign changes occurring
within the
amplitude when $i\gamma_5 \rightarrow 1$, and when $I = 1 \rightarrow I =
0$ [12].  These features allow the expression (28)
to be identified with the instanton contribution to the I = 0 
{\it scalar-channel} sum rule, as well as
with the negative of the corresponding sum rule for the I = 1 channel:

\begin{equation}
\left[ R_0 (\tau) \right]_{inst}^\pi = \left[ R_0 (\tau)
\right]_{inst}^{I = 0} = - \left[ R_0 (\tau) \right]_{inst}^{I = 1} .
\end{equation}

     To summarize, the aggregate field-theoretical contribution from QCD to the leading
Laplace sum-rule (5) is

\begin{eqnarray}
\left[R_0 (\tau , s_0) \right]^{I = 0,1}  & = & \left[ R_0 (\tau, s_0) \right]_{pert}
\nonumber \\
& + & \left[ R_0 (\tau) \right]_{cond} + \left[ R_0 (\tau)
\right]_{inst}^{I = 0,1} ,
\end{eqnarray}

\noindent where the terms on the right-hand side of (32) are respectively given by (22), (27), and
(31) [via
(28)].  Nonperturbative order parameters, specifically the condensates appearing in (27) and the
instanton-size $\rho$ in (28), are known from other theoretical and phenomenological studies -
the
values we quote as standard in Section 4 are consistent with those of refs. 9, 14, and 16.

     As noted earlier on, the next-to-leading sum-rule
$R_1(\tau,s_0)$ can be extracted via
differentiation with respect to the explicit $\tau$ dependence of (32):

\begin{equation}
R_1 (\tau, s_0) = - \frac{\partial}{\partial \tau} R_0 ( \tau , s_0).
\end{equation}

\noindent However, the expressions $R_0(\tau,s_0)$ and
$R_1(\tau,s_0)$ obtained from (32) and (33) require
renormalization-group (RG) improvement to be useful - otherwise results obtained via (9) are
unnaturally dependent on the specific choice for $\alpha_s$.
\footnote{Such is the case in ref. 9, where
$\alpha_s$ is chosen to be 0.6.}  QCD Laplace sum rules have been shown to satisfy RG
equations with
respect to the Borel scale parameter $\tau$ [47].  Consequently, the sum-rules
$R_{0,1}$, once obtained from (32) and (33), can safely be RG-improved by 
replacing $\alpha_s$ with the running coupling-constant
$\alpha_s(\tau^{-1/2})$. In the Section that follows,
factors of $\alpha_s$ in $R_{0,1}$ will be understood to correspond to
the PDG [1] 3-flavour  expression for $\alpha_s(\tau^{-1/2})$ to 3-loop order.

\newpage

\section{Sum Rule Analysis of $q \bar{q}$ Scalar Resonances}

      For a given choice of $\Gamma_\ell, s_0$, and $\tau$, one can obtain a self-consistent value
of the
lowest-lying resonance mass $m_\ell$ by solving (9), as noted in Section 2, thereby defining a
function
$m_\ell(\tau,\Gamma_\ell,s_0)$ implicitly.  Such a procedure is meaningful provided only the
lowest-lying 
$q\bar{q}$
resonance dominates the sum-rule.  This assumption is a reasonable one provided other
subcontinuum resonances in the same channel are either much heavier, and thus exponentially
suppressed (4), or are non-$q\bar{q}$ exotica, as has already been noted. A clear signal of this
assumption's validity is the absence of $a_0$(980), if exotic, from the sum rule generated from
I =
1 $\bar{q}q$-currents (2), even though $a_0$(980) is the lowest lying I = 1 scalar resonance. 
One cannot
explain such an absence if $a_0$(980) is $q\bar{q}$; however its observed absence (see below)
is consistent
with it being both {\it exotic} \footnote{{\it i.e.}., a
$K\bar{K}$-molecule
[2,3], a dynamical-threshold effect [5], or possibly an $s \bar{s}$
state with Zweig-Okubo-suppressed coupling to the nonstrange current
(2).} and {\it sum-rule decoupled}.

      For a given choice of $\Gamma_\ell$ and $s_0$, we shall examine the
$\tau$-dependence of $m_\ell(\tau,\Gamma_\ell,s_0)$
obtained in a given channel from (9) via the field-theoretical expressions for 
$R_{0,1}(\tau,s_0)$ developed
in Section 3. For such externally imposed values of $\Gamma_\ell$ and
$s_0$, we shall identify the true value
of $m_\ell$ with the minimum value of $m_\ell(\tau)$ obtained over an appropriate range of the
Borel
parameter $\tau \equiv 1/M^2$.  Since $M = \tau^{-1/2}$ is itself the renormalization-scale
($\mu$) for the field-theoretical content of $R_{0,1}(\tau,s_0)$, a sum-rule calculation cannot
be meaningful unless we restrict
this mass scale to be not only well-above $\Lambda_{QCD}$, but also to be bounded from above
by the
continuum threshold: $M^2 \leq s_0$.  
This criterion is necessary to ensure that the continuum contribution
to (22), which increases as $s_0 \tau = s_0 / M^2$ becomes small, is
not overly large compared to the remainder of the purely perturbative
contribution, i.e. the $s_0 \rightarrow \infty$ limit of (22).  A
more realistic criterion, in which continuum effects are less than
30\% of the purely perturbative contribution [14], would lead to an
even tighter upper bound on $M$. Moreover, the true value of
$m_\ell$ should not just be the global
minimum of $m_\ell(1/M^2,\Gamma_\ell,s_0)$ 
with respect to M chosen in this subcontinuum range; it should
also exhibit
insensitivity to small changes in the Borel parameter M - a {\it
local} minimum.  This criterion of
flatness, as well as the requirement for a sensible range of the Borel/renormalization scale M, is
quite standard in sum rule applications [40].  We also employ standard values [9,14,16] for
QCD's nonperturbative order parameters in our analysis:
$<m_q\bar{q}q> = -f_\pi^2 m_\pi^2 /4 \; (f_\pi= 131 MeV)$, $<\alpha_s G^2>$ = 0.045
GeV$^2$,$<\alpha_s(\bar{q}q)^2>$= 0.00018 GeV$^6$, $\rho$ = (600
MeV)$^{-1}$.  Factors of $\alpha_s$ that are not absorbed
in (approximately-) RG-invariant QCD-vacuum condensates are replaced with 3-loop
$n_f=3$ running $(\Lambda_{QCD}$ = 150 MeV) coupling constants
$\alpha_s(M) [M \equiv \tau^{-1/2}]$, as discussed at the end of
Section 3.

\subsection{$I = 0$ Scalar Channel: Narrow Resonance Limit}

      In Figure 3, the Borel-scale dependence of the lowest-lying I = 0 scalar resonance mass
(denoted as $M_\sigma$) is displayed for a number of choices of $s_0$, assuming zero resonance
width.  The
curves displayed are restricted to values of $M^2 \leq s_0$.  As in ref. [9], we see that the curves
$M_\sigma(M)$ increase with the choice for $s_0$.  We also see that a subcontinuum local
minimum does
not develop until we consider values of $s_0$ larger than 1.6 GeV$^2$. Figure 4 displays
explicitly the
$s_0$ dependence of the values of $M_\sigma$ that are local minima for each choice of
$s_0$ in Fig. 3. As is
evident from Fig. 4, the minimum value of $M_\sigma$ for a given choice of
$s_0$ also increases with $s_0$. 
Our analysis finds the onset of a local minimum occurring when $s_0$ = 1.61
GeV$^2$, corresponding to $M_\sigma$ = 680 MeV; the Figure 4 curve begins with this point.

      It is useful to compare these results to the seminal 1982 sum-rule analysis of ref. [9],
which included the condensate contributions (27), but which did not include the then-unknown
instanton contribution (28,31) or correct higher-order perturbative contributions.
\footnote{The analysis considered only $O(\alpha_s / \pi)$ 2-loop
contributions to (21), with 17/3 erroneously given as 13/3. The
analysis also did not incorporate RG-improvement of $\alpha_s$.} Figure 5
demonstrates the crucial role instanton contributions play in lowering the lowest-lying 
I = 0 $q\bar{q}$ resonance mass, though at the price of diminishing the broad range of
Borel-parameter stability
observed in [9] at comparable values of $s_0$.  The use of RG-improvement also has a
significant
lowering effect on the lowest-lying I = 0 $q\bar{q}$ resonance mass. In Figure 6, we compare
the
stability curves obtained from the full field-theoretical content of
$R_{0,1}(\tau,s_0)$ at $s_0 = 1.55 GeV^2$
with and without RG-improvement of $[R_{0,1}(\tau,s_0)]_{pert}$.  The upper 
curve is obtained with $\alpha_s$ = 0.6,
as in [9], while the lower curve utilizes $\alpha_s(M)$.  Finally, we note that instanton and RG-
improvement effects are offset somewhat by higher-order perturbative contributions.  Figure 7
demonstrates how going from two to three loops can increase the estimate of the lowest-lying
I = 0 resonance mass by as much as 150 MeV, suggesting higher-order effects as a clear source
of theoretical uncertainty in scalar-channel sum-rule methodology.  Since the leading 4-loop
contribution to $Im[\Pi_{pert}(s)]$ is both large and the same sign
[43] as the leading 3-loop 
$[O(\alpha_s^2)]$ contribution to (21), we can anticipate that sum-rule determinations of Må
based upon (21) will
likely {\it underestimate} $M_\sigma$, a point that will prove important in assessing the viability
of the broad,
very light $\sigma$ (Alternative 1) discussed in Section 1.
\footnote{Our somewhat low value for $\Lambda_{QCD}$ is similarly
motivated to bring down the size of perturbative contributions,
although in practice our results are virtually unaffected by a 100
MeV increase in this parameter.}  

\subsection{$I = 0$ Scalar Channel: Nonzero Widths}

      Figures 8-12 present stability curves analogous to Fig. 3 that are obtained from (9) with
input values of $\Gamma_\ell$ between 100 and 500 MeV.  The weighting functions
$W_{0,1}$ are obtained via
the four-pulse approximation leading to (13) and (14).  For each value of
$\Gamma_\ell$ considered, it is
possible to construct a (Fig. 4 analog) plot of the minimizing value of
$M_\sigma$ for each choice of $s_0$. 
These plots are presented in Fig. 13, and they clearly demonstrate how sum-rule determinations
of the lowest-lying mass increase with increasing resonance width.

      In comparing Figs. 8-12, we note that the onset of a local minimum below the continuum
threshold is itself width-dependent.  Specifically, the $\Gamma$ = 500 MeV stability curves of
Fig. 12
do not develop a locally flat minimum below $s_0$ until $s_0$ is larger than 1.8
GeV$^2$, corresponding
to a minimizing value of $M_\sigma$ at onset that is substantially above 1 GeV.  Local minima
for larger
values of $s_0$ are seen to lead to even larger values of $M_\sigma$, as is evident from the
topmost curve
of Fig. 13.  The values of $s_0$ and $M_\sigma$ corresponding to the onset of local minima are
tabulated
in Table I for $0 \leq \Gamma \leq$ 500 MeV, corresponding to the initial left-hand points of
the six curves
displayed in Fig. 13.  The values of $M_\sigma$ in Table I correspond to the lowest sum-rule
determinations of $M_\sigma$ possible for each choice of $\Gamma$ that are consistent with the
methodological
constraints delineated at the beginning of this Section.  The Table entries clearly indicate that 
{\it a light lowest-lying I = 0  $q\bar{q}$ scalar resonance cannot have a width larger
than half of its mass}.  \footnote{A qualitatively similar conclusion
is stated in ref. [38].}

      This conclusion is softened only slightly if we relax the flatness requirement and insist
only on identifying $M_\sigma$ with the {\it global} minimum over the Borel-parameter
range $(\Lambda_{QCD})^2 < M^2 \leq s_0$.  As $\Gamma$ increases from 100 to 500 MeV,
Figs. 8-12 still show such a global minimum to be
increasing with $\Gamma$.  When $\Gamma$ = 100 MeV, such a minimum occurs at about
530 MeV [see the $s_0 = 1.2 \; GeV^2$ curve of Fig. 8], but as $\Gamma$ increases to 500
MeV, this global absolute minimum increases
to 840 MeV [see the $s_0 = 1.4 \; GeV^2$ curve of Fig. 12].  

      These qualitative conclusions concerning the non-viability of a very
light, very broad $q\bar{q}$ resonance
are not expected to be altered by further refinements in the treatment of the lowest-lying
resonance shape.  Once can obtain ``improved'' weighting-factors
$W_{0,1}$, as opposed to those (13) based upon
a four-pulse approximation of the Breit-Wigner resonance shape
by increasing the value of $n$ used to truncate the Riemann sum (11).
Figure 14 demonstrates that such an increase in the number of pulses
used to approximate the Breit-Wigner shape does not appreciably alter the
local-minimum value $M_\sigma$ obtained for a given choice of $s_0$ and
$\Gamma$.  It should be noted, however,
that $M_\sigma$ increases slightly as $n$ increases, suggesting that a more precise 
modelling of the Breit-Wigner resonance shape would only serve to increase
the $M_\sigma$ values of Table I.

      A more fundamental issue is whether the Breit-Wigner shape is appropriate at all for the
modelling of broad resonances, as the Breit-Wigner tail will extend significantly into Euclidean
$(s < 0)$ and continuum $(s > s_0)$ regions if $\Gamma$ is sufficiently large.  We have relied
upon an
admittedly crude n=4 (4-pulse) approximation in order to minimize such unphysical
contributions.  Larger choices of $n$ increase the sensitivity of $W_{0,1}$ to unphysical regions
in
$s$ by including
square pulses of greater width than the largest pulse in Fig. 2.

      It has already been argued that higher-order perturbative contributions will, if anything,
increase the $M_\sigma$ values of Table I. Consequently, the only means compatible with QCD
sum-rule
methodology that is available for obtaining lower sum-rule estimates for
$M_\sigma$ is to increase the
size of the instanton contributions [{\it c.f.}, Fig. 5], possibly
via a decrease in the value for
$\rho$ in (28). To do so,
however, has ramifications for the I = 1 scalar resonance channel, as discussed below.

\subsection{$I = 1$ Scalar Channel}

      We have seen in Section 3 that the sign of the instanton contribution to
$R_{0,1}$ in the I = 1
scalar channel is reversed from that the I = 0 scalar channel.  Consequently instanton effects are
now seen to increase the scale of sum-rule determinations of the lowest-lying contributing I =
1 resonance.  The stability curves presented in Figs. 15 and 16 clearly indicate a much higher
range of values for masses of such lowest-lying resonances, as well as a need to go to much
higher values of $s_0$ in order to attain subcontinuum stability with respect to the
Borel-parameter
M.

      Table II lists values for the lowest-lying contributing I = 1 resonance mass
($M_\delta$) associated with the onset of a subcontinuum local minimum.  Since local-minima
$M_\delta$ increase
with the choice of $s_0$, the $M_\delta$ values in Table II, like the
$M_\sigma$ values in Table I, represent the
minimum such values obtainable via sum-rule methodology outlined in the beginning of this
Section. Table II clearly indicates a lowest-lying contributing I = 1 scalar resonance in excess
of 1.49 GeV.

      The much larger I = 1 values for the continuum threshold in Table II are comparable to
values for $s_0$ obtained via sum rule analysis of the I = 1 pseudoscalar channel
[18,19].  In both I =
1 channels, sum rule methodology would suggest that Borel stability not occur for values of 
$s_0^{1/2}$ that are less than the masses of contributing subcontinuum resonances. The large
value
of $M_\delta$
evident from Table II necessarily requires values of $s_0$ larger than
$M_\delta^2$, consistent with the values of $s_0$ actually listed.

      As noted at the very beginning of this Section, the results of Table II clearly rule out a
contribution from $a_0$(980), the lowest-lying scalar resonance in the I = 1 channel. 
Consequently,
the coupling of this resonance to the $\bar{q}q$-current (2) [with a minus sign chosen for I =
1] must
be suppressed, as would be anticipated form a non-$q\bar{q}$ interpretation of this state.  The
results of
Table II, however, appear compatible with $a_0$(1450) being identified with the lowest-lying
I = 1
{\it $q\bar{q}$-scalar} resonance, particularly if the width of this resonance is less than or of
order 200 MeV.

      As a final comment, we note that any attempt to obtain sum-rule support for a broad, very
light $q\bar{q}$ state by enhancing the magnitude of the instanton
contribution (e.g. by decreasing the instanton size $\rho$) will necessarily drive up the mass of
the
lowest-lying
contributing I = 1 state, {\it i.e.}, the isovector $q\bar{q}$ state.  
Thus the price of tuning instanton effects to accommodate a light
broad isoscalar $q\bar{q}$ resonance is likely to be an exotic interpretation for both
$a_0$(980) {\it and} $a_0$(1450), the two lowest-lying I=1 scalar resonances.

\newpage

\section{Global-Fit of Lowest-Lying I = 0 Scalar Resonance Properties}

In Section 4, we have utilized a specific choice of the Borel parameter corresponding to
minimization of the sum-rule-derived mass.  However, Leinweber [48] has stressed the value in
an overall fit of the Borel-parameter dependence of a sum rule's field-theoretical content to
the dependence anticipated from resonance properties.  For example, properties of the first
pion-excitation state $\Pi'$ have been obtained by fitting the QCD Borel-parameter dependence
for the lowest $\Pi'$-sensitive Laplace sum rule to its corresponding hadronic content [18,19]. 
In
this section, we will apply this same procedure to the scalar I = 0 channel.  Specifically, we will
obtain a $\chi^2$-minimizing weighted least-squares fit of
$[R_0(\tau,s_0)]^{I=0}$ to the $\tau$-dependence anticipated
from the corresponding resonance contribution (7), assuming the contribution from the lowest-
lying resonance dominates the I=0 scalar channel.

We begin first by modifying the scalar current (2) to include the nonstrange quark mass
in the SU(2)$_f$-invariant limit:

\begin{equation}
j_s(x) = m_q (\bar{u}(x) u (x) + \bar{d} (x) d(x)) / 2
\end{equation}

\noindent where $m_q \equiv (m_u + m_d)/2$.  Although the additional factors of the quark mass
cancel out in an
$R_1/R_0$ determination of resonance properties, such as that in Section 4, these factors have
nontrivial
consequences when the overall $\tau$-dependence of $R_0$ itself is being analyzed.  The current
(34)
now corresponds to an RG-invariant operator, and upon RG-improvement, the sum rules
devolving
from that current (as in Section 3) will now include the additional
$\tau$-dependence associated with the running quark mass:

\begin{eqnarray}
&[R_0(\tau, s_0)]^{I=0}& \nonumber \\
&=&m_q^2 (\tau) \left[ \frac{3}{16 \pi^2 \tau^2} \left\{ \left[ 1 -
(1 + s_0 \tau) e^{-s_0 \tau} \right] \left[ 1 + 17 \alpha_s (\tau) /
3 \pi \right] \right. \right. \nonumber \\
&-& \left. \left. 2(\alpha_s (\tau) / \pi) \int_0^{s_0 \tau} w \; \ell n (w) e^{-w}
dw \right\} \right] \nonumber \\
&+& m_q^2(\tau) \left[ [R_0 (\tau)]_{cond} + [R_0 (\tau)]_{inst}^\pi \right]; 
\end{eqnarray}

\begin{equation}
\alpha_s (\tau) = \frac{2 \pi}{9 \L(\tau)} \left[ 1 - \frac{32}{81 \L
(\tau)} \ell n [\L (\tau)] \right],
\end{equation}

\begin{equation}
m_q (\tau) = \frac{\hat{m}}{[\L(\tau)]^{4/9}} \left[ 1 - \frac{0.1989
- 0.1756 \; \ell n [\L(\tau)]}{\L(\tau)} \right] ,
\end{equation}

\begin{equation}
\L(\tau) \equiv -\frac{1}{2} \ell n \left( \tau \Lambda_{QCD}^2
\right).
\end{equation}

\noindent The condensate and instanton contributions to (35) are given respectively by (27) and
(28).  The
running coupling (36) and mass (37), as well as the purely perturbative contribution to (35), are
given only to two-loop order to accommodate the numerical limitations of our fitting procedure,
which involves generating resonance parameters via a Monte-Carlo simulation of uncertainties
(see below).  The parameter $\hat{m}$ in (37) is the RG-invariant nonstrange quark mass.

Our fit is generated by obtaining values for $m_\sigma, \Gamma_\sigma,
s_0$, and the resonance coupling $g_\sigma$ that
minimize a least-squares fit of (35) to the lowest-lying resonance contribution
anticipated via (7):

\begin{equation}
[R_0(\tau, s_0)]^{I = 0} = g_\sigma W_0 (m_\sigma, \Gamma_\sigma,
\tau) e^{-m_\sigma^2 \tau}.
\end{equation}

In (39), the weighting factor $W_0$ is given by the four-pulse approximation expression (13),
with
$\Delta_0$ given by (15).  Our weighted least-squares fit is over the Borel-parameter range
0.4 GeV$^{-2} \leq \tau \leq$ 2.2 GeV$^{-2}$ [0.67 GeV $\leq$ M $\leq$ 1.6 GeV].  The region
in
$\tau$ is obtained by requiring an
uncertainty of less than 20\% on the theoretical contribution (35) to
$R_0$, based upon a 30\%
continuum and a 50\% power law uncertainty.  Parameter values for perturbative and
nonperturbative scale parameters are as given in Section 4.  As mentioned above, uncertainties
in the fitted parameters $\{m_\sigma, \Gamma_\sigma, s_0, g_\sigma\}$ are obtained only for
the two loop case via a Monte-
Carlo simulation of the power law and continuum uncertainties described above, as well as a
15\%
variation in the size of $\rho$, and a factor of 2 ``vacuum-saturation
uncertainty'' in the value of $<\alpha_s (\bar{q}q)^2>$. 
\footnote{Treatment of this quantity as well as the specific form of
Monte-Carlo modelled uncertainties is discussed at length in ref.
[18].}  The parameter values for $\{m_\sigma, \Gamma_\sigma, s_0,
g_\sigma\}$ we obtain correspond to the fit of (35)
to (39) that minimizes a $\chi^2$ weighted for the previously mentioned continuum and power
law uncertainties.

Figure 17 demonstrates the success of these fitted parameters in matching theoretical (35)
and hadronic (39) expressions for $R_0$.  The results of this fit are as follows:

\renewcommand{\theequation}{40\alph{equation}}
\setcounter{equation}{0}
\begin{equation}
m_\sigma = 0.93 \pm 0.12 GeV,
\end{equation}

\begin{equation}
0 \leq \Gamma_\sigma \leq 0.26 \; GeV,
\end{equation}

\begin{equation}
s_0 = 3.20 \pm 1.20 \; GeV^2,
\end{equation}

\begin{equation}
g_\sigma = \hat{m}^2 \cdot (0.039 \pm 0.014 \; GeV^4),
\end{equation}

\renewcommand{\theequation}{\arabic{equation}}

\setcounter{equation}{40}

\noindent where the quoted uncertainties reflect 90\% confidence levels.  The results (40) are
clearly
indicative of a lowest-lying $q\bar{q}$-resonance that is neither very broad nor very light.  These
results
are most consistent with the Alternatives 2 and 3 delineated in Section 1.

Pertinent to Alternative 3, the results (40) are certainly consistent with identifying the
lowest-lying I = 0 scalar nonstrange $q\bar{q}$ state with $f_0(980)$, as has been suggested by
a very recent
OPAL analysis of $Z^\circ$ decays [6]. Indeed, the very broad region of Borel-parameter stability
characterizing the $s_0 = 2.4$ GeV$^2$ curve of Fig. 3 [which yields an
$f_0(980)$ lowest-lying resonance
mass] corroborates parameter values for $m_\sigma, \Gamma_\sigma$, and
$s_0$ within the fitted range (40).

However, the results (40) do not exclude an Alternative 2 $q\bar{q}$ state somewhat below the
$f_0(980)$ in mass, consistent with a $K\bar{K}$ interpretation of
$f_0(980)$.  Such an $f_0$(980) would be expected to be essentially
decoupled from sum rules based upon the current (34), as has been
discussed in Section 4. This Alternative 2 $q\bar{q}$
interpretation becomes particularly interesting upon examination of the fitted parameter
$g_\sigma$, which
cancels entirely from the $R_1/R_0$-based analysis of Section 4.  Arguments within a
L$\sigma$M context
(buttressed by more general PCAC arguments) have been advanced for
obtaining the matrix
element by which a scalar current couples a physical $\sigma$ to the vacuum [49]:

\begin{equation}
<\sigma|m_q \left(\bar{u}(0)u(0)+\bar{d}(0)d(0) \right)|0> =
f_\pi m_\pi^2 .
\end{equation}
In the above expression, $f_\pi$ is 93 MeV.  In Appendix A we show that 
\begin{equation}
g_\sigma = |<\sigma| j_s (0)|0>|^2 .
\end{equation}
Making use of the I = 0 scalar current $j_s$ defined via (34), we find from
(41) that                                                         
\begin{equation}
g_\sigma = f_\pi^2 m_\pi^4 / 4 .
\end{equation}
Comparison of (40d) and (43) suggests that 
\begin{equation}
\hat{m} = 4.6_{-0.6}^{+1.2} \; MeV ,
\end{equation}
a value somewhat on the low side of the expected range [1] for the nonstrange current quark
mass.  It should be noted, however, that a reasonable light-quark mass
{\it cannot} be obtained in a
dilaton scenario [35,36] in which 

\begin{equation}
<\sigma|\Theta_\mu^\mu (0)|0> = f_\sigma m_\sigma^2 \; \geq \; f_\pi
m_\sigma^2 ,
\end{equation}
unless the contribution to the matrix element from the scalar-current component of
$\Theta_\mu^\mu$ is negligible compared to the anomalous gluonic-field contributions.  If one
assumes that
$\Theta_\mu^\mu \approx m_q(\bar{u}u + \bar{d}d)$, the
quark mass one would obtain by replacing (41) with (45) [via comparison of (40d) with (42)]
will be larger than
(44) by a factor of $m_\sigma^2/m_\pi^2$.

As a final comment, it should be noted that the value for $s_0$ in (40c) is sufficiently high
to
exclude $f_0(1370)$ and $f_0(1500)$ from the continuum.  Although such resonances
(if subcontinuum) are
expected either to be exponentially suppressed [via (7)], or if
non-$q\bar{q}$ exotica, to be essentially
decoupled from a sum rule based on $\bar{q}q$ currents, there is nevertheless reason to be
concerned
about the validity of assuming (as we have done in the section)
that only one resonance contributes to
$R_0$.  To address the possibility of contributions from  more than
one resonance, we have examined whether the field theoretical
content of $[R_0(\tau, s_0)]^{I=0}$ might be better fitted by
contributions from {\it two} subcontinuum resonances by replacing
(39) with
\begin{equation}
R_0(\tau,s_0)]^{I=0} = g_1 W_0(m_1, \Gamma_1, \tau) \exp(-m_1^2
\tau) + g_2  \exp(-m_2^2 \tau).
\end{equation}
The second subcontinuum resonance [presumably $f_0$ (980)] is assumed to be 
narrow: $W_0 (m_2, 0, \tau) =
1$.  The remaining resonance parameters are determined via a weighted
least-squares minimization of $\chi^2$.  We have found such
minimization to be accompanied by equilibration of $m_1$ and $m_2$ to
the previously fitted value (40a), indicating that the
$\tau$-dependence of the field-theoretical content of $R_0$ [Fig. 17]
is best fit by a single exponential (39), rather than by the sum of
two distinct exponential contributions (46).

This result indicates that the leading I=0 scalar-current sum rule is
dominated by a single resonance, as has been assumed in the body of
this section.  Such behaviour is to be contrasted with that of the
leading I=1 pseudoscalar-current sum rule, for which a fit including
two contributing resonances [$\pi$ and $\pi '$] leads to a $\chi^2$
an order of magnitude lower than that of a single resonance fit [18].

\newpage

\section{Conclusions}

      In this paper we have sought to determine which of several empirically justifiable
interpretations of the scalar mesons are compatible with theoretical constraints based upon QCD
sum-rule methodology. 

      For the isovector channel, we have found in Section 4 that
$a_0$(980) is decoupled entirely
from the isovector sum rule based upon the scalar current correlation function given by (1) and
(2), a result requiring an exotic (non-$q\bar{q}$) interpretation for this resonance.  The Table II
values
for the mass of the lowest-lying resonance in this channel that {\it
does} couple to the sum rule appear
compatible with the lowest lying $q\bar{q}$ resonance being $a_0$(1450).

      For the isoscalar channel, we have delineated in the introductory section of this paper four
Alternatives for the lowest-lying  $q\bar{q}$ scalar resonance.  We found in Section 4 that the lowest-lying
I = 0 $q\bar{q}$ scalar resonance cannot be both very light and very broad [Alternative 1], provided this
resonance is the sole contributing resonance to this channel's leading QCD sum rules.  Thus, the
only way the results of Section 4 could accommodate such a resonance would be if this 
lowest-lying resonance's contribution is itself masked by higher-resonance contamination of (7),
the
hadronic side of the sum rule.  The results at the end of Section 5
show this to be unlikely.
Unless we require the Alternative 1 scenario to specify
$f_0$(980) to
be an {\it additional} light $q\bar{q}$ excitation, as opposed to a
$K\bar{K}$ state, and unless the lowest-lying $q\bar{q}$ state is
unnaturally decoupled, such contamination is itself
possible only for values of $s_0$ large enough to include more massive resonances in the
subcontinuum region, such as $f_0$(1370) and $f_0$(1500).  The values of
$s_0$ appearing in Table I are clearly below this threshold.  

      The results of Section 4 show demonstrable support for Alternative 2, a narrower and
more massive lowest-lying resonance that is still lighter than the (presumably exotic)
$f_0$(980),
particularly if the lowest lying resonance is the sole contributing resonance to the sum rule.
Moreover, the results listed in Table I correlate masses for such a resonance with values of
$s_0$
that preclude any contamination of the hadronic side of the sum rule by
$f_0$(1370) and $f_0$(1500). 
These results clearly favour masses above 700 MeV and widths below 300 MeV, and are
suggestive of Svec's most recent single-resonance fits to $\pi N$ data [31].

      It should be emphasized that the results of Section 4 do not exclude the possibility that
$f_0$(980) is itself the lightest I = 0  $q\bar{q}$ scalar
resonance.  Indeed, this
Alternative 3 scenario is strongly supported by the fit performed in Section 5, which gives results
(40) consistent with $f_0$(980) but problematical for a dilaton-explanation of Alternative 2. 
Specifically, the coupling $g_\sigma$ obtained in (40d) is much smaller than that anticipated for
a dilaton,
unless the anomalous gluonic component of $\Theta_\mu^\mu$ dominates the matrix element
(45).  Moreover,
sum-rule contamination from higher I = 0 scalar resonances, though certainly possible for the
fitted range (40c) of $s_0$, would be expected to {\it increase} rather than diminish the apparent
value
of $g_\sigma$ in a single resonance fit, \footnote{Provided the $g_r$ in (7)
are positive-definite: {\it e.g.}, (42).} leading to even a smaller true value for
$g_\sigma$ and an even larger
discrepancy from the scale (45) anticipated from a dilaton interpretation.

      We therefore conclude that the support QCD sum rules provide to Alternative 2 [a lowest
lying resonance below $f_0$(980) whose width is less than half its mass] does
{\it not} extend to a dilaton
interpretation of this state, unless the matrix element (45) is driven almost entirely by
$\Theta_\mu^\mu$'s
anomalous gluon piece.  As a final comment, the Alternative 4 scenario, in which the
lowest-lying $q\bar{q}$ isoscalar
resonance is broad and at least as massive as $f_0$(980), clearly contradicts the QCD
sum
rule results of Section 5. Moreover, even if higher-mass resonance contamination 
of (7) were to occur, such additional hadronic contributions would be
expected to drive the apparent
mass of the lowest-lying state {\it above} its true value, as has already been noted. Thus the
fitted
result (40a) for the lowest-lying resonance mass in the {\it absence} of such contamination cannot
be
reconciled to a true value much in excess of 1 GeV.

\section*{Acknowledgements}

      V. Elias and T. G. Steele would like to acknowledge support from the Natural Sciences and
Engineering Research Council of Canada. V. Elias is grateful for
detailed discussions with V. A. Miransky and M. D. Scadron, as well
as for useful information provided by T. Barnes, G. Lafferty and C.
Strassburger.

\newpage

\appendix
\section*{Appendix A: The $\sigma$ Contribution to the $I = 0$ Scalar
Correlator}

To obtain the explicit $\sigma$ contribution to the I = 0 scalar-current 
$(j_s)$ correlation function (1), we begin by noting that insertion
of a complete set of intermediate single-particle $\sigma$ states
implies that

\renewcommand{\theequation}{A.\arabic{equation}}
\setcounter{equation}{0}
\begin{eqnarray}
& <0| & T j_s (x) j_s (0) |0> \nonumber \\
& = & \Theta (x_0)<0| j_s (x) \left( \int \frac{d^3\vec{q}}{(2 \pi)^3} \;
\frac{|\sigma(\vec{q})><\sigma(\vec{q})|}{2 \sqrt{\vec{q}^2 +
m_\sigma^2}} \right) j_s (0)|0> \nonumber \\
& = & \Theta (-x_0) <0| j_s(0) \left(\int \frac{d^3\vec{q}}{(2 \pi)^3} \;
\frac{|\sigma(\vec{q})><\sigma(\vec{q})|}{2 \sqrt{\vec{q}^2 +
m_\sigma^2}} \right) j_s(x)|0> \nonumber \\
& + & \mbox{(higher-resonance and multiple particle terms)}
\end{eqnarray}

We will define the matrix element connecting a physical (on-shell)
$\sigma$ to the vacuum via a scalar current to be

\begin{equation}
<0| j_s (x) | \sigma (\vec{q})> \equiv M e^{-i(\sqrt{\vec{q}^2+m_\sigma^2}
\; x_0 - \vec{q} \cdot \vec{x})},
\end{equation}
where the Heaviside step function $\Theta(x_0)$ can be expressed as
the following integral:

\begin{equation}
\Theta (x_0) = \lim_{\epsilon \rightarrow 0} \frac{1}{2 \pi i}
\int_{-\infty}^\infty \frac{e^{i x_0 \tau}}{\tau - i | \epsilon |} d
\tau.
\end{equation}

When $x_0 > 0$, the contour along the real $\tau$ axis can be closed
by an infinite arc in the upper half plane, enclosing the residue at
$\tau = i|\epsilon|: \Theta (x_0) = 1$.  When $x_0 < 0$,
the contour must be closed in the lower half plane, and the residue
is not enclosed:  $\Theta(x_0) = 0$.  If we substitute
(A.2) and (A.3) directly into (A.1), we obtain

\begin{eqnarray}
& <0| T j_s (x) j_s (0)|0>_\sigma & \nonumber \\
& = & \lim_{\epsilon \rightarrow 0} \frac{-i |M|^2}{(2 \pi)^4}
\int_{-\infty}^\infty d^3 \vec{q} \int_{-\infty}^\infty d\tau \;
\left[ \frac{e^{-i(\sqrt{\vec{q}^2 + m_\sigma^2} - \tau) x_0} e^{i
\vec{q} \cdot \vec{x}}}{2(\tau - i |\epsilon|) \sqrt{\vec{q}^2 +
m_\sigma^2}}\right. \nonumber \\
& + & \left. \frac{e^{i(\sqrt{\vec{q}^2+m_\sigma^2} - \tau) x_0} e^{-i \vec{q}
\cdot \vec{x}}}{2(\tau - i|\epsilon|) \sqrt{\vec{q}^2 + m_\sigma^2}}
\right]
\end{eqnarray}
where the $\sigma$ subscript on the left-hand side of (A.4)
represents the explicit contribution of single-$\sigma$ intermediate
states to the matrix element.  To evaluate (A.4) further, we make the
following change of variable from $\tau$ to $q_0$:

\begin{equation}
q_0 \equiv \sqrt{\vec{q}^2 + m_\sigma^2} - \tau \; \; , \; \; d q_0 =
-d \tau.
\end{equation}

\noindent We then obtain $(d^4 q \equiv d^3 \vec{q} \; d q_0 \; ; \; q
\cdot x \equiv q_0 x_0 - \vec{q} \cdot \vec{x})$:

\begin{eqnarray}
&<0|&T j_s (x) j_s (0) |0>_\sigma \nonumber \\
& = & \lim_{\epsilon \rightarrow 0} - \frac{i|M|^2}{(2 \pi)^4} \int
_{-\infty}^\infty d^4 q \left[ \frac{e^{-i q \cdot x}}{2 (
\sqrt{\vec{q}^2 + m_\sigma^2} - q_0 - i|\epsilon|) \sqrt{\vec{q}^2 +
m_\sigma^2}} \right. \nonumber \\
& + & \left. \frac{e^{i q \cdot x}}{2(\sqrt{\vec{q}^2 + m_\sigma^2} -
q_0 - i |\epsilon|) \sqrt{\vec{q}^2 + m_\sigma^2}}\right] \nonumber
\\
&=& \lim_{\epsilon \rightarrow 0} - \frac{i|M|^2}{(2 \pi)^4}
\int_{-\infty}^\infty d^4 q \frac{e^{-i q \cdot x}}{2 \sqrt{\vec{q}^2
+ m_\sigma^2}} \left[ \frac{1}{\sqrt{\vec{q}^2 + m_\sigma^2} - q_0 -
i|\epsilon|} \right. \nonumber \\
&+& \left. \frac{1}{\sqrt{\vec{q}^2 + m_\sigma^2} + q_0 - i|\epsilon|}
\right]
\end{eqnarray}
The final line of (A.6) is obtained from the intermediate line by
letting $q_\mu \rightarrow -q_\mu$ in the second portion of the
integral.  We find trivially from (A.6) that $(q^2 \equiv q_o^2 -
\vec{q}^2 \; ; \; \epsilon' = 2 \sqrt{\vec{q}^2 + m_\sigma^2} \;
\epsilon)$;

\begin{eqnarray}
& <0|T j_s (x) j_s(0)|0>_\sigma & \nonumber \\
&=& \lim_{\epsilon' \rightarrow 0} + \frac{i|M|^2}{(2 \pi)^4}
\int_{-\infty}^\infty d^4 q \; \frac{e^{-i q \cdot x}}{q^2 - m_\sigma^2
+ i |\epsilon'|} .
\end{eqnarray}
Upon substitution of (A.7) into (1), we find that

\begin{equation}
(\Pi(p^2))_\sigma = - \frac{|<\sigma| j_s (0)|0>|^2}{p^2 - m_\sigma^2
+ i|\epsilon'|}
\end{equation}
where we have utilized the explicit definition of the constant $M$ in
(A.2).  The propagator denominator in (A.8) can be modified
in the standard way [1] for incorporating relativistically
a Breit-Wigner width $\Gamma_\sigma$:
\begin{equation}
(\Pi(p^2))_\sigma = - \frac{|<\sigma| j_s (0)|0>|^2}{p^2 - m_\sigma^2
+ i m_\sigma \Gamma_\sigma}.
\end{equation}
Consequently, we see from comparison to (3) that the coupling coefficient
$g_r$ is given by
\begin{equation}
g_r = |<\sigma| j_s (0)|0>|^2
\end{equation}

\newpage

\begin{center}
\begin{tabular}{lclllll}
$\Gamma$(MeV): & \underline{0} & \underline{100} & \underline{200} &
\underline{300} & \underline{400} & \underline{500} \\
$s_0$(GeV): & 1.61 & 1.61 & 1.62 & 1.65 & 1.70 & 1.82 \\
$m_\sigma$(MeV): & 680 & 687 & 716 & 778 & 884 & 1087
\end{tabular}
\end{center}
\smallskip
\baselineskip=12pt
{\bf Table I:} Sigma masses associated with the onset of a
subcontinuum local minimum in the isoscalar channel.

\bigskip

\bigskip

\begin{center}
\begin{tabular}{lclll}
$\Gamma$(MeV): & \underline{0} & \underline{100} & \underline{200} &
\underline{300} \\
$s_0$(GeV$^2$): & 3.24 & 3.26 & 3.38 & 3.62 \\
$M_\delta$(GeV): & 1.49 & 1.50 & 1.55 & 1.63
\end{tabular}
\end{center}
\smallskip
{\bf Table II:} Lowest-lying masses $(M_\delta)$ associated with the
onset of a subcontinuum local minimum in the isovector channel.

\newpage

\section*{Figure Captions:}

\begin{description}
\item {\bf Figure 1:} An example of the 4 square-pulse approximation to the Breit-Wigner
resonance
shape obtained by truncating eq. (11) to n=4, and by choosing f = 0.701 to ensure
that the area under the four pulses is equivalent to the total area under the Breit-Wigner curve. 
This particular example is for a mass M = 680 MeV and width
$\Gamma$ = 100 MeV.
\end{description}

\begin{description}
\item[Figure 2:] \item{(a) The single-instanton contribution to the quark-antiquark scattering
amplitude
           ({\it c.f.} fig. 5 of ref. [12]).}
\item{(b)The single-instanton contribution to the corresponding current-current
           correlation function.}
\end{description}

\begin{description}
\item {\bf Figure 3:} Stability curves for determining via (9) the lowest-lying I = 0 resonance
mass. The
corresponding width of this resonance is assumed to be zero.  The stability curves
are truncated to exclude the region in which the Borel mass-scale M exceeds
$s_0^{1/2}$.
\end{description}

\begin{description}
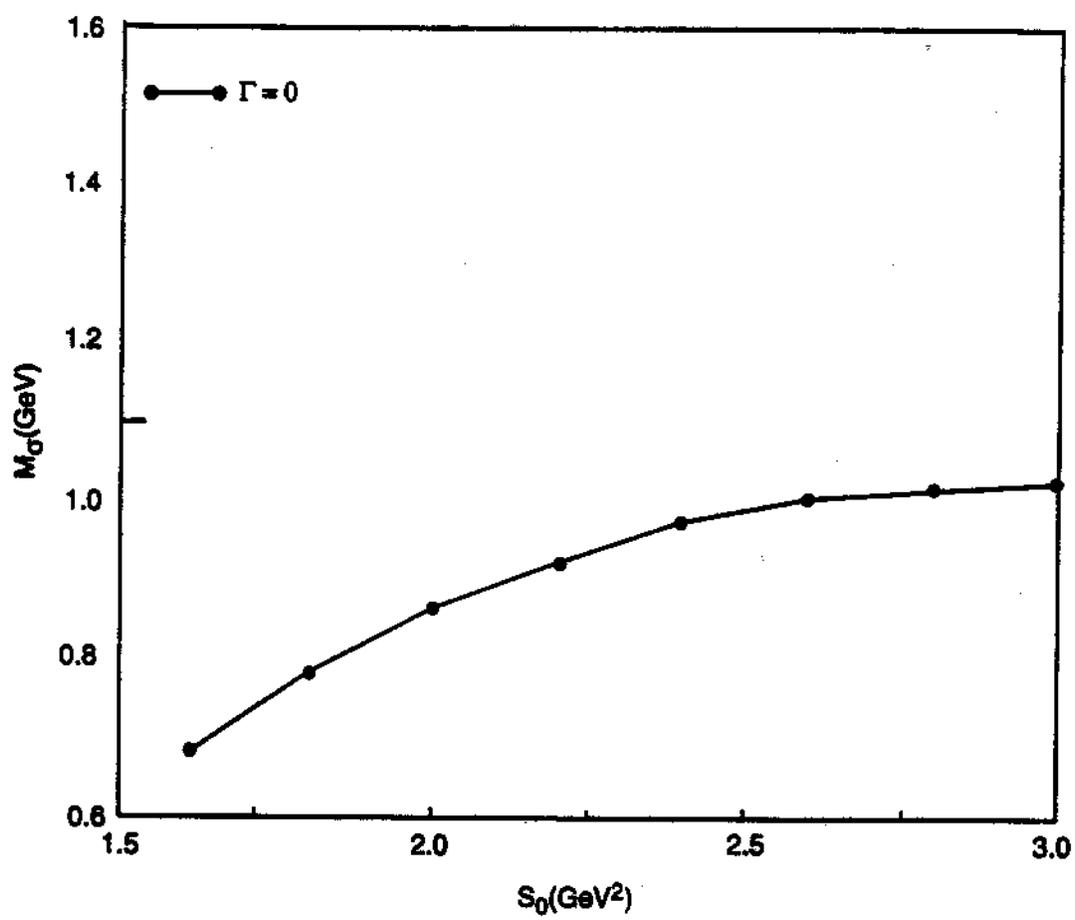
\item{\bf Figure 4:} The $s_0$-dependence of the masses $M_\sigma$ corresponding to (local)
minima of the
           stability curves presented in Fig. 3.
\end{description}

\begin{description}
\item{\bf Figure 5:} How instanton contributions affect $M_\sigma$: a comparison of two
stability curves for
           the lowest-lying I = 0 resonance mass obtained by including or not
           including the direct single-instanton contributions to the field-theoretical
           expressions for $R_{0,1}(\tau,s_0)$.  Both curves are obtained assuming zero resonance
           width and a continuum threshold $s_0$ = 1.55 GeV$^2$.
\end{description}

\begin{description}
\item{\bf Figure 6:} How renormalization-group improvement affects
$M_\sigma$: a comparison of two stability
curves for the lowest-lying I = 0 resonance are obtained by including or 
not including the $\tau$ dependence of the strong coupling.  The constant coupling
curve is obtained by choosing $\alpha_s$ = 0.6, as in ref. [9].  The (3-loop) running
coupling curve is obtained assuming $\Lambda_{QCD}$ = 150 MeV.  Both curves are obtained
assuming zero resonance width and a continuum threshold $s_0$ = 1.55
GeV$^2$, and both curves incorporate direct single-instanton contributions 
to the field-theoretical expressions for $R_{0,1}(\tau,s_0)$.
\end{description}

\begin{description}
\item{\bf Figure 7:} How three-loop perturbative contributions affect
$M_\sigma$: a comparison of two stability
curves for the lowest-lying I = 0 resonance mass obtained by including
or not including three-loop order perturbative contributions to the field-theoretical
expressions for $R_{0,1}(\tau,s_0)$. As in Figs. 5 and 6, $s_o$ is assumed here to be 1.55
GeV$^2$, and the resonance width $\Gamma$ is assumed to be zero.
\end{description}

\begin{description}
\item{\bf Figure 8:} Stability curves for determining via (9) the lowest-lying I = 0 resonance
mass,
assuming a resonance width of 100 MeV. As in Fig. 3, the stability curves are
truncated to exclude the region in which the Borel mass scale M exceeds
$s_0^{1/2}$.
\end{description}

\begin{description}
\item{\bf Figure 9:} Stability curves for determining the lowest-lying I = 0 resonance mass,
assuming
           a resonance width of 200 MeV.
\end{description}

\begin{description}
\item{\bf Figure 10:} Stability curves for determining the lowest-lying I = 0 resonance mass,
assuming
           a resonance width of 300 MeV.
\end{description}

\begin{description}
\item{\bf Figure 11:} Stability curves for determining the lowest-lying I = 0 resonance mass,
assuming
           a resonance width of 400 MeV.
\end{description}

\begin{description}
\item{\bf Figure 12:} Stability curves for determining the lowest-lying I = 0 resonance mass,
assuming
           a resonance width of 500 MeV.
\end{description}

\begin{description}
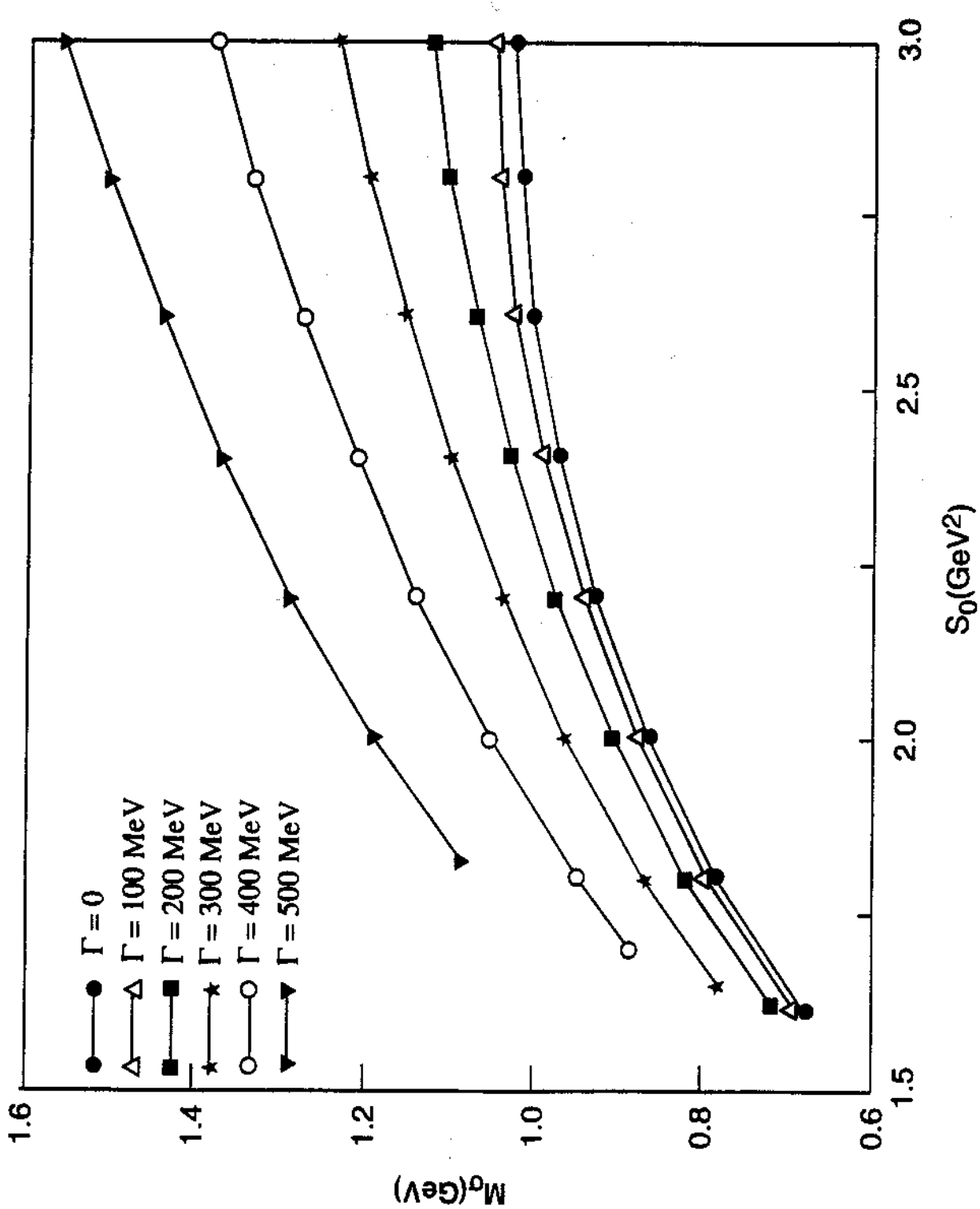
\item{\bf Figure 13:} The $s_0$-dependence of the masses $M_\sigma$ corresponding to (local)
minima of the
           stability curves presented in Figs. 8-12.
\end{description}

\begin{description}
\item{ \bf Figure 14:} How improving the 4 square-pulse approximation affects
$M_\sigma$: a comparison of
three stability curves for the lowest-lying I = 0 resonance mass that are obtained
via (9) from weighting functions $W_{0,1}(M_\sigma,\Gamma,\tau)$ based upon truncation of
the
Riemann sum (11) at n = 4, 8, and 14 pulses.  A larger number of pulses
corresponds to a more precise approximation of the Breit-Wigner shape than that
presented in Fig. 1. In all three curves, $s_0$ is assumed to be 1.63
GeV$^2$, and the resonance width $\Gamma$ is assumed to be 100 MeV.
\end{description}

\begin{description}
\item{\bf Figure 15:} Stability curves for determining via (9) the lowest-lying
$I = 1$ resonance mass,
denoted here as $M_\delta$. The corresponding width of this resonance is assumed to be
zero.  As before, the stability curves are truncated to exclude the region in which
the Borel mass scale $M$ exceeds $s_0^{1/2}$.
\end{description}

\begin{description}
\item{\bf Figure 16:} Stability curves for determining the lowest-lying I = 1 resonance mass,
assuming
           a resonance width of 200 MeV.
\end{description}

\begin{description}
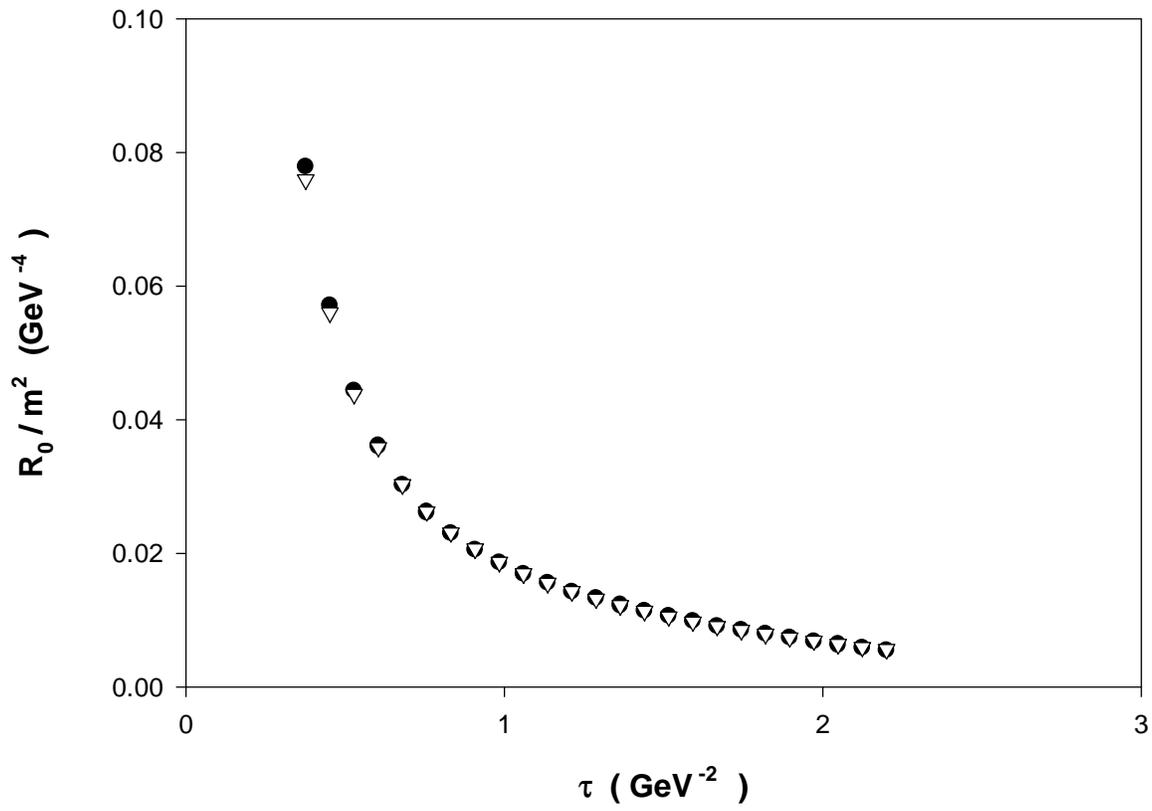
\item{\bf Figure 17:} Weighted least-squares fit of field theoretical content
(circles) of
$R_0(\tau,s_0)/ \hat{m}^2$ to the
corresponding hadronic contribution (triangles) anticipated from the lowest-lying 
$I=0$ scalar resonance. Parameter values leading to this fit are
$m_\sigma$ = 0.95 GeV, $\Gamma_\sigma = 0$, $s_0$ = 3.3 GeV$^2$, and
$g_\sigma/\hat{m}^2$ = 0.040 GeV$^4$.
\end{description}
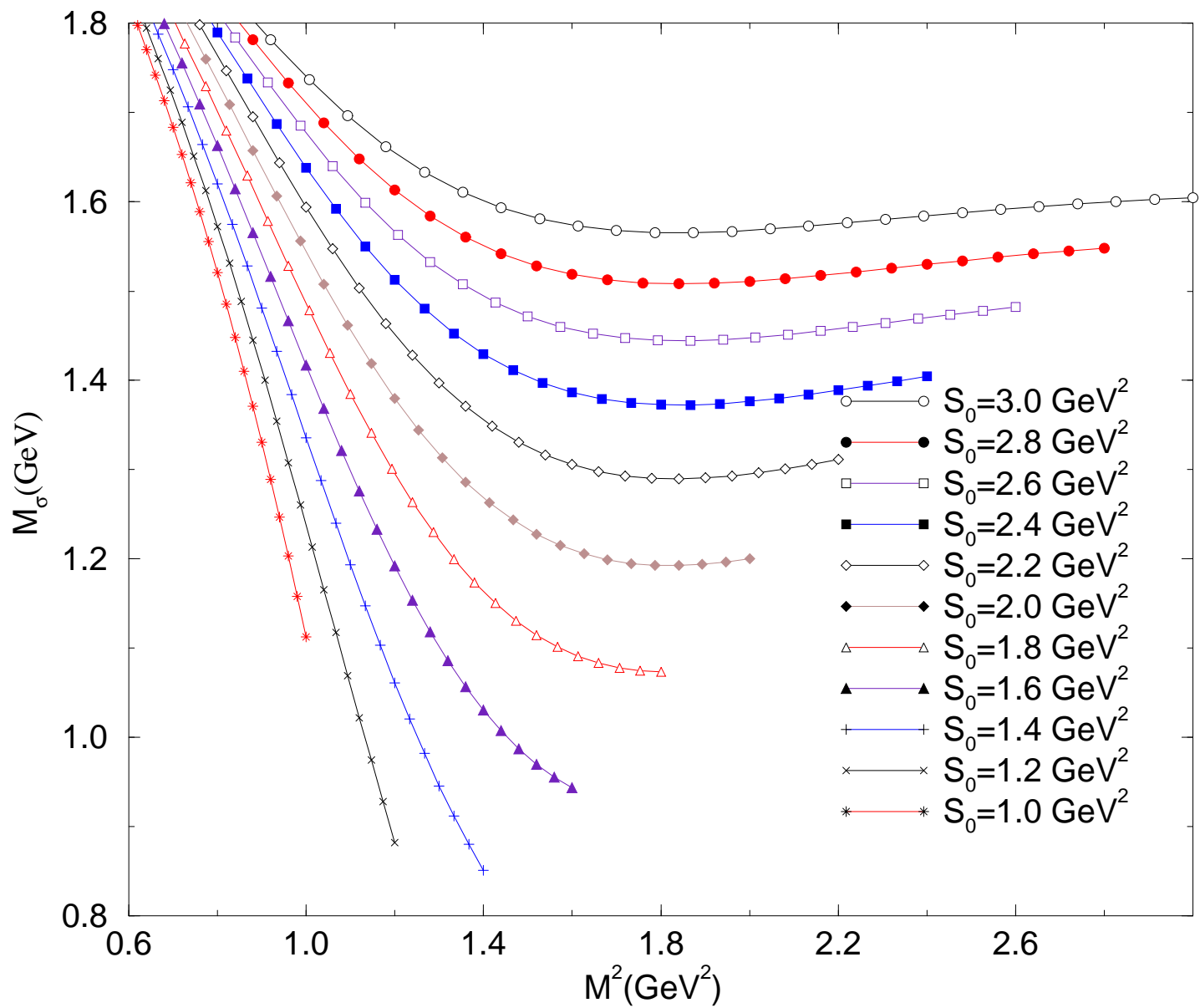
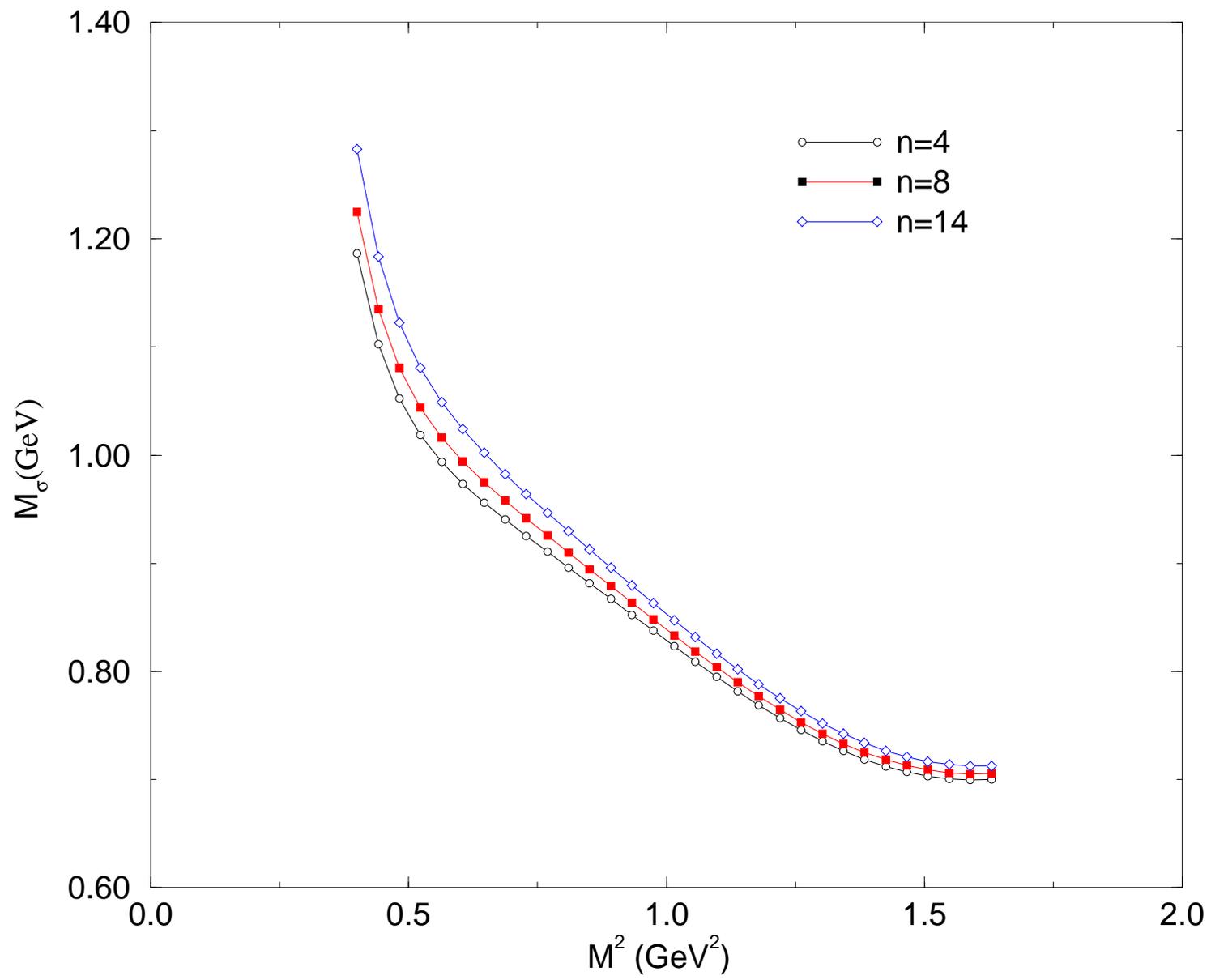
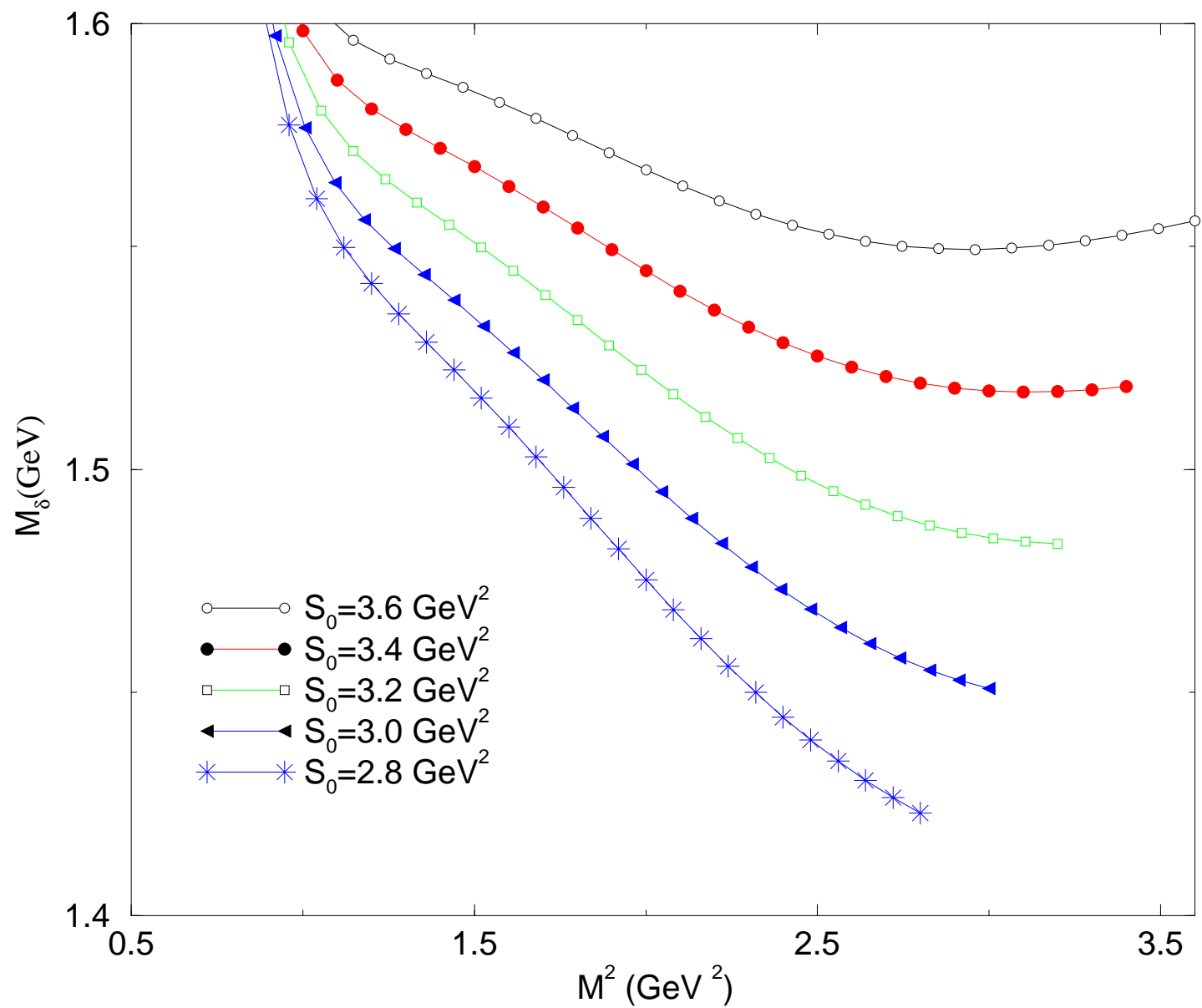
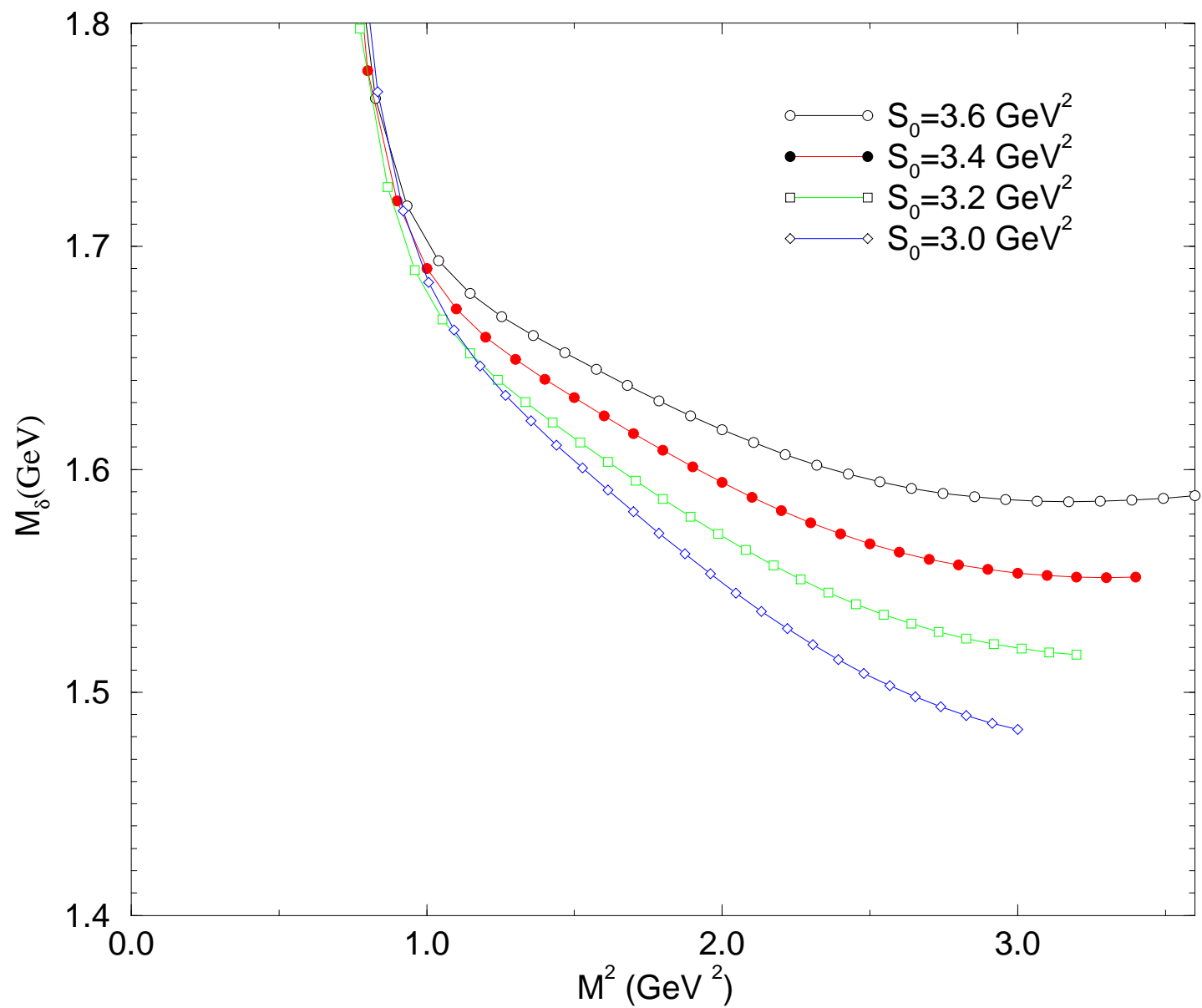
\end{document}